\documentclass[pdflatex,sn-mathphys-num]{sn-jnl}

\usepackage{graphicx}%
\usepackage{multirow}%
\usepackage{slashed}
\usepackage{amsmath,amssymb,amsfonts}%
\usepackage{amsthm}%
\usepackage{mathrsfs}%
\usepackage[title]{appendix}%
\usepackage{xcolor}%
\usepackage{textcomp}%
\usepackage{manyfoot}%
\usepackage{booktabs}%
\usepackage{algorithm}%
\usepackage{algorithmicx}%
\usepackage{algpseudocode}%
\usepackage{listings}%

\theoremstyle{thmstyleone}%
\theoremstyle{thmstyletwo}%

\theoremstyle{thmstylethree}%

\begin{document}
\title[Article Title]{$\pi$- and $K$-Mesons  Properties for  Large $N_f$}


\author{\fnm{Aftab} \sur{Ahmad}}\email{aftabahmad@gu.edu.pk}



\affil*{\orgdiv{Institute of Physics}, \orgname{Gomal University}, \orgaddress{\street{} \city{D.I.Khan}, \postcode{29022}, \state{Khyber Pakhtunkhaw}, \country{Pakistan}}}




\abstract{The restoration of chiral symmetry and deconfinement for a higher number of light-quark flavors $N_f$ implies suppression of the dynamically generated quark mass. The study of various larger values of $N_f$  may have a greater impact on the internal structure of light hadrons.  In this work, we study the properties of the $\pi$- (pion) and $K$-meson (kaon), such as the mass, in-meson condensate and leptonic decay constant, for various $N_f$.  In this context, we employ the symmetry-preserving, confining vector–vector flavor-dependent contact interaction (FCI) as an input to the Schwinger–Dyson equation (SDE) and the homogeneous Bethe–Salpeter equation (BSE). In the chiral limit where $(m_f = 0)$, increasing the number of flavors $N_f$ leads to the restoration of dynamical chiral symmetry and  deconfinement at a critical number of flavors,  $N^{c}_{f} \approx 8$.  At this threshold, all parameters vanish, and the confinement scale $\tilde{\tau}_{ir}$ diverges. Below this critical number of flavors $N_f < N^{c}_{f}$, the mass of the Goldstone boson $m_{GB}$ (or massless pion) remains unchanged, indicating that  chiral symmetry is still broken. However, for $N_{f} > N^{c}_{f}$, where chiral symmetry is fully restored, $m_{GB}$ increases rapidly. This transition signifies that the Goldstone boson shifts from being a bound state to a resonant particle, eventually dissociating at a critical number of flavors $N^{d}_{f}$, which coincides with $N^{c}_{f}$. This dissociation of bound states intuitively signals deconfinement, as quarks and antiquarks can be released from their bound states. On the other hand, when the bare quark mass for the up and down
$m_{u,d}=0.007$, and the strange bare quark mass $m_s=0.17$ GeV  is considered,
the dynamically generated quark mass $M_{u/d}$, strange quark mass $M_s$,  in-meson condensate $\kappa^{1/3}_{(\pi,K)}$,  and decay constant $f_{(\pi, K)}$ 
monotonically decrease as $N_f$ increases. As in this scenario, the chiral symmetry explicitly broken and partially restored at  and above $N^{d}_{f} \approx 8.2$. The  masses of pion and kaon  $m_{(\pi, K)}$  increases  quickly and dissociate at and above  critical dissociation number of flavors $N^{d}_{f} =N^{d}_{(\pi,K)f} \approx 8.2$, i.e., this $N^{d}_{(\pi,k)f}$ separate the bound states of the pion and kaon from  their constituents, similar to their behavior  at Mott temperature in the presence of a heat bath. 
We also verified the consistency of our findings across different flavors using the Gell-Mann-Oakes-Renner relation.
}

\keywords{Dynamical chiral symmetry breaking, confinement, chiral Symmetry breaking, Confinement, Schwinger-Dyson Eqautions, Bethe-Salpeter Eqaution, light hadrons}



\maketitle

Quantum Chromodynamics (QCD) is a well-established theory that explores the fascinating world of strong color interaction between quarks and gluons. These interaction formed the bound states known as hadrons.  QCD has two major regimes known as asymptotic freedom and quark confinement. Asymptotic freedom, which was discovered by Gross and Wilczek
~\cite{Gross:1973id, Politzer:1973fx}, describes the weak interaction between quarks at high energies. On the other hand, quark confinement, as proposed by Wilson~\cite{Wilson:1974sk}, explains the strong interaction between quarks at low energies, preventing them from existing in isolation. In addition to the color confinement, 
the dynamical chiral symmetry breaking is another significant characteristic of low-energy QCD ~\cite{Nambu:1961fr, Nambu:1961tp,Maskawa:1974vs,Atkinson:1986ay}. This concept is closely linked to the generation of  constituent quark masses ~\cite{Praschifka:1987ry}.
It is widely recognized that QCD displays both confinement and chiral symmetry breaking when considering a small number of light quark flavors, denoted as $N_f$~\cite{Engel:2014cka,Evans:2020ztq,Ciambriello:2024xzd}. However, for larger values of $N_f$, it is believed that there exists a critical value $N_{f}^{c}$, beyond which the chiral symmetry is restored and the quark becomes deconfined~\cite{Appelquist:2009ty,bashir2013qcd,LSD:2014nmn}. This critical value $N_{f}^{c}$ must be smaller than the upper limit of the  critical value, denoted as $N_f^{AF,c}$, the threshold beyond which asymptotic freedom is lost. According to the research by Politzer in 1973~\cite{Politzer:1973fx}, for a gauge group $SU(3)$, the critical number is  $N^{AF,c}_{f}=16.5$. Therefore, the QCD theory is considered to be conformal in the infrared regime, driven by the presence of an infrared fixed point (the fixed point corresponds to a specific point where the $\beta$-functions for the QCD couplings tends to  zero)~\cite{Caswell:1974gg,Banks:1981nn,gies2006chiral,Appelquist:2007hu,hasenfratz2010conformal,aoki2012many,Evans:2020ztq} .  The range of fermion flavors $N^{c}_{f}\lesssim N_f<N^{AF,c}_{f}$, it is commonly referred to as the ``conformal region"~\cite{Appelquist:2009ty, LSD:2009yru}. As we approach  the upper limit $(N_f \lesssim N^{A,c}_{f})$ of this region, the infrared fixed point is situated in the weakly interacting region. However, at the lower end $(N_f \sim N^{c}_{f})$, the infrared fixed point undergoes a shift towards the strongly interacting region. In this region, the coupling becomes increasingly strong as $N_f$ decreases. Consequently, the system enters a phase characterized by the breaking of chiral symmetry and the confinement of quarks.
Lattice QCD simulations~\cite{LSD:2014nmn,Hayakawa:2010yn,Cheng:2013eu,Hasenfratz:2016dou,LatticeStrongDynamics:2018hun}, as well as continuum methods of QCD~\cite{bashir2013qcd,Appelquist:1999hr,Hopfer:2014zna,Doff:2016jzk,Binosi:2016xxu,Ahmad:2020jzn} and some effective models of QCD~\cite{Ahmad:2022hbu} in the fundamental $SU(3)$ representation have all highlighted the significance of the chiral symmetry restoration and deconfinement phases occurring as $N_f$ enters the conformal zone $8\lesssim N^{c}_{f}< 12$ and approximated the value of critical number of flavors $N^{c}_{f}\approx 8$.
In recent comprehensive analysis~\cite{Ahmad:2022hbu, ahmad2025qcdphasediagramlarge}, the QCD phase diagram at finite temperature and density has been explored, with a particular focus on the influence of higher $N_f$. The findings indicate that the critical line separating the hadron phase from the quark gluon plasma experiences suppression as the number of flavors increases. Additionally, the Schwinger quark-antiquark production rate  exhibits a swifter growth  with the electric field strength as the number of flavors, $N_f$, increases,~\cite{Ahmad:2023mqg}.\\
Exploring the impact of higher values of  $N_f$ on light hadron bound states, especially the properties of the $\pi$- and $K$- meson, would yield intriguing and meaningful results. The $\pi$- and $K$-meson  are considered the lightest hadrons and serves as mediators for the long-range interaction between hadrons. These mesons are easily generated in collisions involving electrons and nucleons, making them ideal for investigating models of hadronic structure and sub-nucleonic degrees of freedom within nuclei. Being mesons, they consist of a simple quark-antiquark valence-quark composition, making them the simplest light-quark systems to study as bound states influenced by strong interactions. There is a strong motivation in electron-ion-collider (EIC) in BNL  beside HERA to probe into the structure of $\pi$- and $K$-meson, see for example Ref.~\cite{Aguilar:2019teb} for detail analysis .
Our primary aim and motivation for this endeavor and research undertaking is to explore various parameters associated with light hadrons, particularly the masses, in-pseudoscalar meson condensate and leptonic decay constant of $\pi$- and $K$-meson, when considering a larger number of quark flavors $N_f$. In this context, we can trigger the critical number of flavor $N^{d}_{f}$ for  bound state dissociation in analogy to the Mott temperature in QCD at finite temperature, where the dissociation of bound states (hadrons) into their constituent quarks and gluons, as temperature increases~\cite{Hufner:1996pq,Contrera:2009hk,Xu:2021lxa}. It will be worth to explore the bound state dissociation number of flavors $N^{d}_{f}$ for the bound state of pseudoscalar mesons in the chiral limit $m_f=0$,  as well as with the bare quarks masses ($m_f\neq0$) which have not been explore so for to the best of our knowledge.
This investigation holds great importance as it allows us to gain a deeper understanding of how these fundamental particles behave under varying $N_f$.\\ 
In our present manuscript, we employ the Schwinger-Dyson equations to calculate the dressed quark mass, and we utilize the homogeneous Bethe-Salpeter equation~\cite{Salpeter:1951sz} to determine the mass of the $\pi$- and $K$- meson for higher values of  $N_f$. The homogeneous BSE  can be understood as an eigenvalue problem, where the mass of the bound
state can be determined from an eigenvalue
satisfying the following identity $\lambda( P^{2}= -m_{H}^{2}) = 1$, where
$P$ is the momentum and $m_H$ is the mass of the bound state meson~\cite{Bedolla:2015mpa,Maris:1997tm}. The eigenvector in this case corresponds to the  bound state amplitude. This bound state amplitude, also known as the Bethe-Salpeter amplitude (BSA), plays a vital role in the computation of production and scattering processes that involve mesons. By delving into these calculations, we aim to gain a deeper understanding of the intricacies and characteristics of light hadrons for large $N_f$.
Our gap equation and the BSE kernel will be formulated using the symmetry-preserving, confining vector-vector contact interaction~~\cite{GutierrezGuerrero:2010md, Roberts:2011cf, Roberts:2011wy}, whose $N_f$-dependent has been introduced in Ref.~\cite{Ahmad:2020jzn, Ahmad:2022hbu,Ahmad:2023mqg}. We will be working in the Landau gauge and employing an optimal Schwinger-proper time regularization scheme within the rainbow-ladder truncation. This approach allows us to maintain the desired symmetries while accurately capturing a sensible constituent-like value for the quark mass in the infrared in our calculations. 
The article is organized as follows: In Section II, we provide a comprehensive overview of the flavor-dependent contact interaction model and the SDE gap equation. In Section III,  we present the BSE approach to mesons. Additionally, we discuss the meson bound state quark interaction model, including the FCI model approach and the  BSA for mesons. Moving on to Section IV , we present our desired findings and results, focusing on a higher number of flavors. Lastly, in Section V, we summarize our research and highlight the outcomes we have achieved. 
\section{Flavor-dependent  Contact Interaction Model and the Gap equation} \label{section-2}
 
We start our discussion from the Schwinger-Dyson equations ~\cite{Schwinger:1951nm}, where the dressed-quark propagator $S_f$ can be written as:
\begin{eqnarray}
S^{-1}_{f}(p)&=S^{-1}_{f,0}(p) + \Sigma(p)\,.\label{CI1}
\end{eqnarray}
Here, $S_{f,0}(p)=(\slashed{p}+ m_f + i\epsilon)^{-1}$, is the bare quark propagator.  The $S_{f}(p)=(\slashed{p}+ M_{f} + i\epsilon)^{-1}$ represents the dressed quark propagator. Here $M_f$ is the dress quark mass and  $m_f$ is bare quark mass. Where $\Sigma(p)$ is the self-energy and is given by:
\begin{eqnarray}
\Sigma(p)=\int \frac{d^4q}{(2\pi)^4} g^{2}
 D_{\mu\nu}(p-q)\frac{\lambda^a}{2}\gamma_\mu S_{f}(q)
\frac{\lambda^a}{2}\Gamma_\nu(p,q)\,,\label{CI2}
\end{eqnarray}

Where $g^2$ is the running coupling associated with the quark-gluon vertex and $\Gamma_\nu$ is the dressed quark-gluon vertex,  with $p$ and $q$ representing the four-momenta of the quark and $(p-q)$ denoting the four-momentum of the gluon.  Here, $\gamma_\mu (\gamma_\nu)$ are the Dirac gamma matrices in a $4$-dimensional space. The $D_{\mu\nu}$ is the gluon propagator and 
 $\lambda^a$'s are the Gell-Mann matrices with $a=1,..8$, represents the color indices. In the ${\rm SU(N_c)}$ representation the Gell-Mann matrices satisfy the following identity :
\begin{eqnarray}
\sum^{N_{c}^{2}-1}_{a=1}\frac{{\lambda}^a}{2}\frac{{\lambda}^a}{2}=\frac{1}{2}\left(N_c - \frac{1}{N_c} \right)I, \label{CI3}
\end{eqnarray} 
where, $I$ is the identity matrix and $N_c$ represents the number of colors.\\
In the  symmetry-preserving contact interaction ~\cite{GutierrezGuerrero:2010md,Roberts:2011cf,Roberts:2011wy,Qin:2011dd,Chen:2012qr}, $\Gamma_\nu =\gamma_\nu $, the gluon propagator (in the Landau gauge) along with the running coupling (see, for example, Ref.~\cite{Deur_2024}  for detail and references in) are considered in the deep infrared region  where the gluons acquire a dynamically generated running mass $m_{g}$~\cite{Langfeld:1996rn, Cornwall:1981zr,PhysRevD.80.085018, Aguilar:2015bud,Oliveira:2010xc,Boucaud:2010gr, Kohyama:2016obc,Ferreira:2025anh}, are assumed to be momentum independent~\cite{GutierrezGuerrero:2010md,Roberts:2011cf,Roberts:2011wy,Chen:2012qr, Xing:2021dwe}. This  momentum-independent  contact interaction  can effectively describe the static properties of light pseudoscalar and vector mesons~\cite{GutierrezGuerrero:2010md,Roberts:2011cf,Roberts:2011wy,Chen:2012qr, Xing:2021dwe}, yielding results comparable to those derived from more advanced QCD model interactions~\cite{Maris:2006ea,Bashir:2012fs,Eichmann:2008ae,Cloet:2007pi}. Here, we follow the assumption from Ref.~\cite{Chen:2012qr,Bedolla:2015mpa}  as follows: 
\begin{eqnarray}
  g^2 D_{\mu\nu }(p-q)\rightarrow\frac{4 \pi
  \alpha_{\rm ir}} {m_{g}^{2}} \delta_{\mu \nu} \doteq \delta_{\mu \nu} \alpha_{\rm eff}   \,.\label{CI4}
\end{eqnarray}
This ensures the mapping of one-gluon exchange diagrams to a contact interaction. The fitted parameter $\alpha_{\rm ir}=0.93\pi$ represents the infrared-enhanced interaction strength, aligning with current estimates of the zero-momentum value of the running coupling in QCD~\cite{Maris:2005tt, Aguilar:2010gm, Brodsky:2010px,Deur:2016tte, Aguilar:2015bud, Deur_2024}. The gluon mass scale $m_g = 0.8$ GeV is typically associated with the one-loop renormalization-group-improved interaction detailed in~\cite{Qin:2011dd, Chen:2012qr, Bedolla:2015mpa}. The values of the parameters mentioned above have been fine-tuned to reproduce the properties of light hadrons~\cite{GutierrezGuerrero:2010md, Roberts:2011cf, Roberts:2011wy,Qin:2011dd, Chen:2012qr, Bedolla:2015mpa}. It should also be noted that $4\pi\alpha_{ir}/m^{2}_{g}$ of~\cite{Qin:2011dd, Chen:2012qr}$= 1/m^{2}_{G}$ (with $m_G = 0.132$ GeV) of~\cite{GutierrezGuerrero:2010md, Roberts:2011cf, Roberts:2011wy} (see, for example, Eq.(13) of \cite{Wang:2013wk} or Eq.(5) of~\cite{Bedolla:2015mpa}). With a specific choice of the gap equation kernel, the dressed quark mass function is simply a constant.\\
To study chiral symmetry breaking and restoration for large $N_f$ using symmetry-preserving contact interactions,~\cite{Ahmad:2020jzn, Ahmad:2022hbu, ahmad2025qcdphasediagramlarge} modified the effective coupling $\alpha_{\rm eff}$ in a way that provides solutions to the gap equation for a higher number of flavors. With this modification, one can determine the critical number of flavors $ N^{c}_{f}$ for chiral symmetry breaking or restoration. The motivation for this modification is based on arguments presented in~\cite{bashir2013qcd, Ahmad:2020jzn, Ahmad:2022hbu, ahmad2025qcdphasediagramlarge}, where the critical number of flavors $N^{c}_{f}$ obtained from Schwinger-Dyson equations with two different truncations has been utilized. According to~\cite{bashir2013qcd, Ahmad:2020jzn, Ahmad:2022hbu, ahmad2025qcdphasediagramlarge}, the dynamically generated mass $M_f$ should exhibit the following relationship with the critical number of flavors $ N^{c}_{f}$:
\begin{eqnarray}
M_f \backsim \sqrt{1 - \frac{N_f}{N^{c}_{f}}}.\label{CI4a}
\end{eqnarray}
 To achieve this behavior in the four-fermion contact interaction, only a square root $N_f$-dependence in the coupling results in the observed behavior~\cite{Ahmad:2020jzn, Ahmad:2022hbu, ahmad2025qcdphasediagramlarge}:  
\begin{eqnarray}
\alpha_{\rm eff}(N_f)= \alpha_{\rm eff}\sqrt{1 - \frac{(N_{f}-2)}{\mathcal{N}_{f}^{c}}} \,.\label{CI4b}
\end{eqnarray}
Here, $\alpha_{\rm eff}(N_f)$ is the flavor-dressed effective coupling for the contact interaction. For a fixed number of flavors, $N_f = 2$, this four-fermion contact interaction effective coupling $\alpha_{\rm eff}(N_f)$ reduces to the CI effective coupling $ \alpha_{\rm eff}$. The term $\mathcal{N}^{f}_{c} = N^{c}_{f} + \varsigma$ in Eq.~(\ref{CI4b}) represents estimated values for the critical number of flavors ~\cite{Ahmad:2020jzn, Ahmad:2022hbu}. The parameter $\varsigma=2.9$, is adjusted in this work  achieve the desired critical number $N^{c}_{f} \approx 8$, above which dynamical chiral symmetry is restored and deconfinement occurs. Ref.~\cite{Ahmad:2020jzn} explains and justifies the introduction of the parameter $\varsigma$ as resulting from the factor $(N_f - 2)$ in Eq.~(\ref{CI4b}). With a particular choice of $N_f-$flavor dependent effective coupling Eq.~(\ref{CI4b}), we have from Eq.~(\ref{CI4}), the FCI model is given by: 
\begin{eqnarray}
  g^2 D_{\mu\nu }(p-q)\rightarrow \delta_{\mu \nu} \alpha_{\rm eff}(N_f)   \,.\label{CI4c}
\end{eqnarray}
With a particular choice of the flavor-dependent model Eq.~(\ref{CI4c}),  using $N_c=3$ and after performing the trace over the Dirac matrices, the dynamical quark mass function  $M_f$ is given by:
\begin{equation}
M_f=m_f+\frac{4\alpha_{\rm eff}(N_f)}{3}\int \frac{d^{4}q}{(2\pi)^4} {\rm Tr}[S_{f}(k)]\;.\label{CI6}
\end{equation}

On simplifying Eq.~(\ref{CI6}), we get 
\begin{eqnarray}
M_f=m_f+\frac{16\alpha_{\rm eff}(N_f)}{3}\int \frac{d^4q}{(2\pi)^4} \frac{M_f}{q^2+M^{2}_{f}}\;.\label{CI7}
\end{eqnarray}
After setting the variable $s=q^2$, and using  $d^4 q= (1/2) q^2 dq^2 \sin ^2 \theta  d\theta \sin \phi d \phi d\psi $ in Eq.~(\ref{CI7}) and then performing the integration, we get
\begin{eqnarray}
M_f=m_{f}+\frac{M_f\alpha_{\rm eff}(N_f)}{3\pi^2}\int^{\infty }_{0}
ds\frac{s}{s+M_f^2} \,. \label{CI9}
\end{eqnarray}
The integral mentioned in equation  Eq.~(\ref{CI9}) is not convergent and requires regularization. In this study, we utilize the Schwinger proper-time regularization scheme~\cite{Schwinger:1951nm} to address this issue. This procedure involves exponentiation the denominator of the integrand and introducing an additional infrared cut-off $\tau_{ir}=1/\Lambda_{ir}$, alongside the conventional ultraviolet cutoff $\tau_{uv}=1/\Lambda_{uv}$ . The integrand of Eq.~(\ref{CI9}) can be written as
\begin{eqnarray}
\frac{1}{s+M^{2}_{f}}&=&\int^{\infty }_{0} d\tau {\rm e}^{-\tau(s+M^{2}_{f})}
\rightarrow
 \int^{\tau_{ir}^2}_{\tau_{uv}^2} d\tau
 {\rm e}^{-\tau(s+M^{2}_{f})} \nonumber\\ &=&
\frac{ {\rm e}^{-\tau_{uv}^2(s+M^{2}_{f})}-{\rm e}^{-\tau_{ir}^2(s+M^{2}_{f})}}{s+M^{2}_{f}}\;. \label{CI10}
\end{eqnarray}
The infrared cutoff $\tau_{ir}$ is employed to implement confinement by ensuring the absence of quark production thresholds~\cite{Ebert:1996vx, Roberts:2007jh, Roberts:2011cf, Roberts:2011wy}. However, it is important to note that the CI model, as described in Eq.~(\ref{CI9}), is a non-renormalizable theory. This means that the parameter $\tau_{uv}$, cannot be eliminated and instead becomes an integral part of the model. It sets the scale for all dimensional quantities within the model. Furthermore, the ultraviolet cutoff $\Lambda_{uv}$, also plays a significant role in the study of heavier quarks. By increasing $\Lambda_{uv}$, we can simulate the short-distance effects that occur as the quark mass increases.   The adoption of this regularization procedure is significant as it eliminates quadratic and logarithmic divergences and ensures that the axial-vector Ward-Takahashi identity is satisfied~\cite{Ward:1950xp, Takahashi:1957xn}. From  Eq.~(\ref{CI9}) and Eq.~(\ref{CI10}) and  after performing the  integration over `s', the gap equation is reduced to:
\begin{eqnarray}
 M_f&=& m_{f} + \frac{M_f \alpha_{\rm eff}(N_f)}{3\pi^{2}} \mathcal{A}_{01}(M_f^2,\tau_{\rm uv}^2,\tau_{\rm ir}^2)
 ,\label{CI11}
\end{eqnarray} 
with 
\begin{equation}
\mathcal{A}_{\delta \zeta}(M^2;\tau_{\rm uv}^{2},\tau_{\rm ir}^{2} )= \frac{(M^{2}_{f})^{\epsilon}}{\Gamma (\zeta)} \Gamma(\zeta-2,\tau_{\rm uv}^{2} M^2, \tau_{\rm ir}^{2} M^2),\label{CI12}
\end{equation}
\noindent where $\epsilon =\delta-(\zeta-2)$ and 
$
\Gamma (a, y_1,y_2)=\Gamma (a,y_1)-\Gamma(a,y_2)$
with $\Gamma(a,y) = \int_{y}^{\infty} t^{\alpha-1} {\rm e}^{-t} dt$ being the incomplete Gamma function.
In this FCI model, the confinement  for higher $N_c$ and $N_f$ can be triggered from the  confining length scale\cite{Ahmad:2016iez, Ahmad:2020ifp, Ahmad:2020jzn,Ahmad:2023mqg,Ahmad:2023ecw,Ahmad:2025prr,ahmad2025qcdphasediagramlarge}:
\begin{eqnarray}
\tilde{\tau}_{ir}=\tau_{ir}\frac{M(2)}{M(N_f)}.\label{CI13}
\end{eqnarray}
 Here $M(2)$ is the dressed mass for $N_f=2$.  $M(N_f)$  is the generalized $N_f$ dependent dressed mass. In the presence of  $N_f$,  $\tilde{\tau}_{ir}$ needs to vary slightly with $N_f$. Therefore, the connection between dynamical chiral symmetry breaking and confinement is expressed through a regulator, denoted as $\tilde{\tau}_{ir}$, which explicitly depends on $N_f$ . The confinement length scale, as described in Eq.~(\ref{CI13}), in the chiral limit  becomes infinite when chiral symmetry is restored (see inside the Fig.~\ref{Fig1} in the numerical section). This leads to deconfinement at and above a certain critical value $N_{f}^{c}$, where the production thresholds reappears~\cite{Ahmad:2020jzn,Ahmad:2023mqg,ahmad2025qcdphasediagramlarge}, is consistent with the choice of Eq.~(\ref{CI10}) and behave like NJL interaction.  In the next section, we discuss the general formalism for BSE with FCI Model.

\section{Bethe-Salpeter Equation for Pseudoscalar  and  the  FCI model}
A comprehensive relativistic approach to meson bound states is facilitated by the BSE. The BSE reveals poles in the four-point function, which correspond to the existence of meson-bound states. The homogeneous BSE establishes the conditions necessary for these poles to emerge in a specific $J^{PC}$ channel. Since quarks cannot be directly observed in high-energy experiments, the study of bound states becomes crucial for testing QCD. Specifically, the determination of a meson bound-state problem in a specific channel $J^{PC}$ is based on its homogeneous counterpart and is given by:
\begin{equation}
\Gamma_{H}(k;P)=
\int\frac{d^{4}q}{(2\pi)^4} K^{(2)}(k,q;P)\chi(q;P),\label{BSE1}
\end{equation}
with
\begin{equation}
\chi(q;P)=S_{f}(q_{+})\Gamma_{H}(q;P)S_{g}(q_{-}), \label{BSE1a}
\end{equation}
is the Bethe-Salpeter wave-function and  $q_{+}=q+\eta P$,
$q_{-}=q-(1-\eta)P$; $\eta \in [0,1]$ is a momentum-sharing parameter. Here, $k$ is the relative momentum and $P$ is the
total momentum of the quark-antiquark system. $\Gamma_{H}(k;P)$ represents the Bethe-Salpeter meson amplitude. The symbol $H=f\bar{g}$ denotes the quantum numbers and the flavor composition of the meson.  In the equations above, we omit all renormalization constants, as well as the indices related to color, flavor, and spin, for the sake of simplicity in notation. As previously stated, $S_{f}$ is the dressed-quark propagator and  $K^{(2)}(k,q;P)$ is the two body irreducible quark-antiquark scattering kernel. The study of the  meson  observables lacks significance unless they explicitly ensure the satisfaction of the vector and axial-vector Ward-Takahashi identities (axWTI)~\cite{Maris:1997hd}. In the chiral limit ( $m_f = 0 $), axWTI can be written as:
\begin{equation}
- P_{\mu}\Gamma_{5\mu}(k;P)=S^{-1}(k_{+})\gamma_{5} + \gamma_{5}S^{-1}(k_{-}).\label{BSE2}
\end{equation}
Where $k_{+} = k + P $, $k_{-} = k $  and  $\Gamma_{5\mu}(k,P)$ is the axial vector vertex, which is determined by
\begin{equation}
\Gamma_{5\mu}(k;P)=\gamma_{5}\gamma_{\mu}+\int\frac{d^{4}q}{(2\pi)^4}  K^{(2)}(k,q;P) \chi_{5\mu}(q_{+},q)\\. \label{BSE2a}
\end{equation}
Eq.~(\ref{BSE2}), which capture the phenomenological aspects of dynamical chiral symmetry breaking in QCD, it is essential to establish a connection between the axial-vector vertex, denoted as $\Gamma_{5\mu}(k;P)$, and the quark propagator, represented as $S_{f}(k)$. This connection implies a correlation between the kernel $K^{(2)}(p,q;P)$ in the BSE  Eq.~(\ref{BSE1}) and the one-body kernel $K^{(1)}(p,q;P)$  in the quark SDE Eq.~(\ref{CI2})~\cite{Xing:2021dwe}. Any truncation scheme applied to the SDE-BSE coupled system must preserve this relationship, thereby influencing the characteristics of the $K^{(2)}(p,q;P)$.
In the context of the contact interaction and rainbow-ladder truncation ~\cite{Maris:1997hd,Roberts:2010rn,Roberts:2011cf}, the $K^{(2)}(p,q;P)$ is given as \cite{Xing:2021dwe}:
\begin{equation}
   K^{(2)}(p,q;P)= -\frac{4}{3}g^{2}D_{\mu\nu}(p-q)
  \gamma_{\mu}\otimes
  \gamma_{\nu}= - K^{(1)}(p,q;P),\label{BSE3}
\end{equation}
Here, $K^{(1)}$  is  related to
the ‘rainbow’ piece, while $K^{(2)}$ corresponds  to the ‘ladder ’ part, of the rainbow-ladder kernals. Inserting the  FCI model Eq.~(\ref{CI4c}) in Eq.~(\ref{BSE3}), we have
\begin{equation}
   K^{(2)}(p,q;P)= -\frac{4}{3}\alpha_{\rm eff}(N_f)\delta_{\mu\nu}
  \gamma_{\mu}\otimes
  \gamma_{\nu},\label{BSE3a}
\end{equation}
and then using in Eq.~(\ref{BSE1}),  we have the homogeneous BSE ($\eta=1$) for the ground state meson is given by
\begin{eqnarray}
 \Gamma_{H}(k;P) = -\frac{4}{3}\alpha_{\rm eff}(N_f)\int \frac{d^{4}q}{(2\pi)^{4}} \nonumber\\ 
 \gamma_{\mu}S_{f}(q+P)\Gamma_{H}(q;P)S_{\bar{g}}(q)\gamma_{\mu}. \label{BSE3b}   
\end{eqnarray}
Eq.~(\ref{BSE3b}), for the ground state pseudoscalar mesons $(J^{PC}=0^{-+})$, can be expressed as:
\begin{eqnarray}
 \Gamma_{0^{-+}}(k;P) = -\frac{4}{3}\alpha_{\rm eff}(N_f)\int \frac{d^{4}q}{(2\pi)^{4}} \nonumber\\ 
 \gamma_{\mu}S_{f}(q+P)\Gamma_{0^{-+}}(q;P)S_{\bar{g}}(q)\gamma_{\mu}. \label{BSE4}   
\end{eqnarray}
In the symmetry-preserving regularization scheme and in the rainbow-ladder truncation the BSA cannot depend on
relative momentum (see for example the detailed analysis~\cite{Roberts:2010rn, Roberts:2011cf}).
A general decomposition for
pseudoscalar meson in the FCI  scenario has the following form: 
\begin{equation}
\Gamma_{0^{-+}}(P) = i\gamma_{5}E_{0^{-+}}(P)+\frac{1}{M_{R}}\gamma_{5}\gamma.PF_{0^{-+}}(P), \label{BSE5}
\end{equation}
where $M_R=M_f M_{\bar{g}}/(M_f+M_{\bar{g}})$, $E_{0^{-+}}(P)$ and $F_{0^{-+}}(P)$ are the Bethe-Salpeter amplitudes (BSAs). 
Substituting  Eq.~(\ref{BSE5}) in Eq.~(\ref{BSE4}),  and by adopting the same procedure described in Refs.~\cite{Roberts:2010rn,Roberts:2011cf,Bedolla:2015mpa,Raya:2017ggu,Gutierrez-Guerrero:2019uwa} allows for the easy derivation of the explicit expression for the BSE of the pseudoscalar mesons. To ensure the preservation of Eq.~(\ref{BSE2}), it is necessary to incorporate a regularization technique. This necessity leads to the following form of axWTI~\cite{Roberts:2010rn,Roberts:2011cf,Bedolla:2015mpa,Raya:2017ggu,Gutierrez-Guerrero:2019uwa}:  
\begin{equation}
0=  \int^{1}_{0}d\alpha   \int  \frac{d^{4}q}{(2\pi)^{4}} \frac{\frac{1}{2}q^{2}+\mathcal{W}}{[q^{2}+\mathcal{W}]^{2}}, \label{BSE5a}
\end{equation}
where $\alpha$  is a Feynman parameter and $\mathcal{W}=
\mathcal{W}(M_f,M_{\overline{g}},\alpha, P^2)=M_{f}^{2}(1-\alpha)+ M_{\overline{g}}^{2}\alpha+\alpha(1-\alpha)P^{2}$. Eq.~(\ref{BSE5a}) can be further simplified as 
\begin{equation}
0=\int^{1}_{0}d\alpha \lbrace{\mathcal{A}_{01}(\mathcal{W})+\mathcal{A}_{02}(\mathcal{W})}\rbrace\,,\label{BSE5b}
\end{equation}
with $\mathcal{A}_{02}(x)= -x(d/dx)\mathcal{A}_{01}(x)$
 and  $\bar{\mathcal{A}}_{02}(x)=\mathcal{A}_{02}(x)/x$.
Now we can write the explicit form of the Bethe-Salpeter equation Eq.~(\ref{BSE4})  in the FCI model as:
\begin{equation}
     \begin{pmatrix}
     E_{0^{-+}}(P)\\
     F_{0^{-+}}(P)\\ 
\end{pmatrix}
=\frac{\alpha_{\rm eff}(N_f)}{3\pi^{2}}
\begin{pmatrix}
    \mathcal{ K}^{0^{-+}}_{EE}   \mathcal{K}^{0^{-+}}_{EF}\\
    \mathcal{ K}^{0^{-+}}_{FE}   \mathcal{K}^{0^{-+}}_{FF}\\ 
\end{pmatrix}
\begin{pmatrix}
    E_{0^{-+}}(P)\\
     F_{0^{-+}}(P)\ 
\end{pmatrix}.\label{BSE6}
\end{equation}
Eq.~(\ref{BSE6}), is an Eigenvalue equation which  has a solution at $P^2=-m^{2}_{0^{-+}}$, with $m_{0^{-+}}$ is the mass of pseudoscalar meson.  The explicit form of the kernels of the BSE Eq.~(\ref{BSE6})  for the pseudoscalar meson (See for example for detail, Appendix A of Ref.~\cite{Bedolla:2015mpa} and  Ref.~\cite{Gutierrez-Guerrero:2019uwa}) are:

\begin{eqnarray}
    \mathcal{K}^{0^{-+}}_{EE}=\int^{1}_{0}d \alpha \left\lbrace \mathcal{A}_{01}(\mathcal{W})
- [M_f M_{\overline{g}}\alpha(1-\alpha)P^{2}-\mathcal{W}]\bar{\mathcal{A}}_{02}(\mathcal{W})\right\rbrace,\nonumber
\end{eqnarray}

\begin{equation}
\mathcal{K}^{0^{-+}}_{EF}=\frac{P^2}{2M_{R}} \int^{1}_{0}d{\alpha}[M_{f}(1-\alpha)+\alpha M_{\overline{g}}]\bar{\mathcal{A}}_{02}(\mathcal{W}),\nonumber
\end{equation}
\begin{equation}
\mathcal{K}^{0^{-+}}_{FF}=-\frac{1}{2}  \int^{1}_{0}d{\alpha}[M_f M_{\overline{g}}+M^{2}_{f}(1-\alpha)+M^{2}_{\overline{g}}\alpha]\bar{\mathcal{A}}_{02}(\mathcal{W}),\nonumber
\end{equation}
\begin{equation}
\mathcal{ K}^{0^{-+}}_{FE}= \frac{2M^{2}_{R}}{P^2}\mathcal{K}^{0^{-+}}_{EF}.\label{BSE7}
\end{equation}
For the purpose of computation of the 
observable,  the BSA has
to be normalized. In the rainbow ladder
truncation of the BSE, the normalization condition is of the form:
\begin{eqnarray}
P_{\mu}=&&N_c\int\frac{d^{4}q}{(2\pi)^{4}} Tr[\Gamma_{0^{-+}}(-Q)\nonumber\\
&&\times\frac{\partial}{\partial P_\mu}  S_{f}(q+P)\Gamma_{0^{-+}}(Q)S_{\bar{g}}(q)],\label{BSE7a}
\end{eqnarray}
where $N_{c}=3$. The condition Eq.~(\ref{BSE7a}) can be further simplified as:
\begin{eqnarray}
1=&& \frac{d}{dP^2} 2N_{c}\int\frac{d^{4}q}{(2\pi)^{4}} \nonumber\\
&&\times Tr[\Gamma_{0^{-+}}(-Q) S_{f}(q+P)\Gamma_{0^{-+}}(Q)S_{\bar{g}}(q)],\label{BSE7b}
\end{eqnarray}
with $P^2 = -m^{2}_{0^{-+}}$, where $m_{0^{-+}}$ represents the mass of pseudoscalar meson and at $P=Q$. By utilizing the amplitudes and propagators, it is possible to calculate various characteristics of the pseudoscalar mesons (the $\pi$-meson  and $K$-meson  in the present case) within the rainbow ladder truncation. This includes the leptonic decay constants  and the  condensates for the mesons. The decay constant in terms of canonically normalized amplitudes can be expressed as~\cite{Maris:1997hd,Maris:1997tm, Roberts:2010rn,Brodsky:2012ku,Chang:2011mu}):
\begin{equation}
f_{0^{-+}}
P_{\mu} =N_c \int\frac{d^{4}q}{(2\pi)^{4}} Tr[\gamma_{5}\gamma_{\mu}S(q_{+})\Gamma_{0^{-+}}(P)S(q_{-})],\label{BSE8}
\end{equation}
or on simplifying, we have (see for example~\cite{Chen:2012txa}):
\begin{equation}
f_{0^{-+}}=\frac{1}{M_R}\frac{N_{c} }{4\pi^{2}}[E_{0^{-+} }\mathcal{ K}^{0^{-+}}_{FE}+F_{0^{-+}}\mathcal{ K}^{0^{-+}}_{FF}],\label{BSE8a}
\end{equation}
and the in-meson condensate  is defined in Refs.~\cite{Maris:1997hd,Maris:1997tm}, see also~\cite{Brodsky:2012ku,Chang:2011mu, Chen:2012txa}, for the ground state  pseudoscalar meson~\cite{Brodsky:2010xf}, is expressed as:
\begin{equation}
\kappa_{0^{-+}} = (f_{0^{-+}})N_c\int\frac{d^{4}q}{(2\pi)^{4}} Tr[\gamma_{5}S(q_{+})\Gamma_{0^{-+}}(P)S(q_{-})]\,.\label{BSE8b}
\end{equation}
The simplified version of the Eq.~(\ref{BSE8b})~\cite{Roberts:2010rn,Chen:2012txa} can be written as:
\begin{equation}
\kappa_{0^{-+}}=
f_{0^{-+}}\frac{1}{M_R}\frac{N_{c} }{4\pi^{2}}[E_{0^{-+}}\mathcal{ K}^{0^{-+}}_{EE}+F_{0^{-+}}\mathcal{ K}^{0^{-+}}_{EF}].\label{BSE8c}
\end{equation}
Moreover, in the chiral limit $m_{f,g}\rightarrow 0$ , the in-pseudoscalar meson condensate\cite{Brodsky:2010xf} implies the so-called vacuum quark condensate~\cite{Maris:1997tm, Maris:1997hd}:
\begin{equation}
\lim_{m_{f,g}\to0} \kappa_{0^{-+}}
= N_c\int\frac{d^{4}q}{(2\pi)^{4}} Tr[S(q)]=-\langle \bar{q}q\rangle^0, \label{BSE8d}
\end{equation}
where the superscript``0'' is stand for chiral limit. The vacuum quark condensate is, in fact, the
chiral-limit value of the in-pseudoscalar meson condensate ( i.e., it describes
a property of the chiral-limit pion ) ~\cite{Brodsky:2010xf}. It would be beneficial to express the relationship among the leptonic decay constant, the ground state in-pseudoscalar meson condensate, and the mass of the pseudoscalar meson using the Gell-Mann-Oakes-Renner relation~\cite{Gell-Mann:1968hlm}, as discussed in~\cite{Maris:1997tm, Maris:1997hd,Brodsky:2012ku, Roberts:2010rn, Chang:2011mu,Chen:2012txa}
\begin{equation}
f_{0^{-+}}^{2} m_{0^{-+}}^{2} =\frac{1}{2} [m_{f} +m_{g}] \kappa_{0^{-+}}.\label{BSE8e}
\end{equation}
 In the chiral limit it has been verified the Gell--Mann--Oakes--Renner relation for the pseudoscalar meson Eq.~(\ref{BSE8e}) i.e., in the neighborhood of the chiral limit~\cite{Maris:1997hd,Chang:2011mu}, is given by:
\begin{equation}
m_{0^{-+}}^2 = -[m_f+m_g] \frac{\langle \bar q q \rangle^{0}}{(f^{0}_{0^{-+}})^{2}} + {\rm O}(m_{fg}^2).\label{BSE8f}
\end{equation}
With $f^{0}_{0^{-+}}$, is the pseudoscalar meson decay constant in the chiral limit. Here in this particular manuscript, we are studying the properties of $\pi$- and $K$-meson, we can re-write the GMOR relations Eq.~(\ref{BSE8e}) as:
\begin{equation}
f_{\pi}^{2} m_{\pi}^{2} =\frac{1}{2} [m_{u} +m_{d}] \kappa_{\pi}.\label{BSE8g}
\end{equation}
and 
\begin{equation}
f_{\pi}^{2} m_{K}^{2} =\frac{1}{2} [m_{u} +m_{s}] \kappa_{K}.\label{BSE8h}
\end{equation}

In the next section, we discuss the numerical solution of the SDE-BSE  to calculate the properties of $\pi$ -meson and $K$ -meson for a higher number of light-quark flavors $N_f$ from the FCI model.

\section{Numerical results} \label{section-3}

\begin{figure}
\includegraphics[width=8cm]{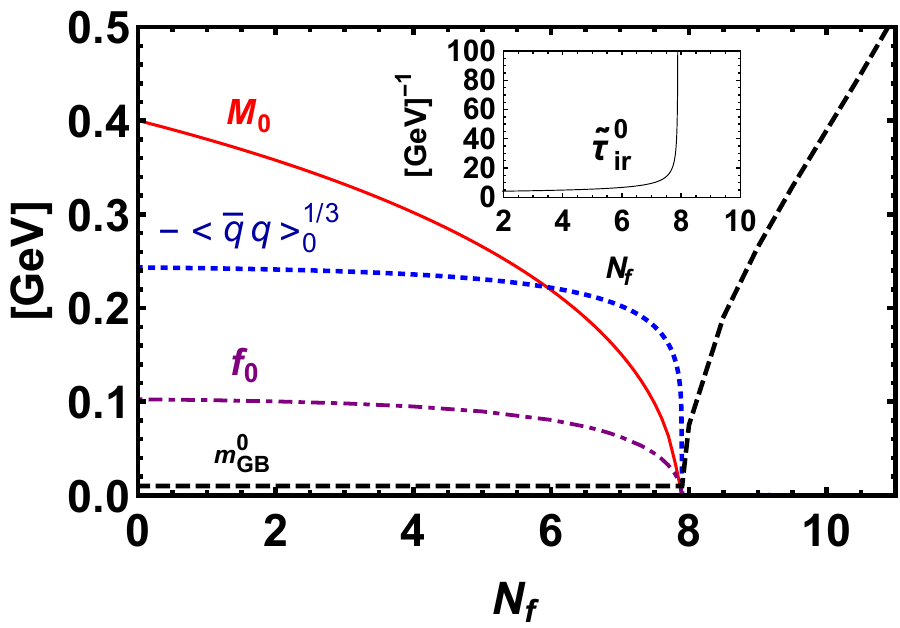}
\caption{ In the chiral limit, behavior  of the dressed quark mass $M_0$, the masses of the Goldstone boson $m^{0}_{GB}$ (massless pions) , the leptonic decay constant $f_{0}$, the quark-antiquark condensate $-<\bar{q}q>^{1/3}$, and the confinement scale $\tilde{\tau}$ for various number of flavors $N_f$. All parameters, except for the confinement scale and the massless bosons, vanish when $N_f$ reaches or exceeds $N^{c}_{f}\approx8$, indicating the restoration of dynamical chiral symmetry. The confinement scale $\tilde{\tau}$ diverges, signaling a transition to deconfinement. In the chiral limit, $N_f$-dependence of dressed quark mass $M_0$, Goldstone boson (massless pion) masses $m^{0}_{GB}$, leptonic decay constant $f_{0}$, the quark-antiquark condensate ($-<\bar{q}q>^{1/3}$) and the confinement scale $\tilde{\tau}$ small figure inside. All the parameters  except the confinement scale and massless boson, vanishes at and above $N^{c}_{f}\approx8$, where the dynamical chiral symmetry is restored. The confinement scale $\tilde{\tau}^{0}$ diverges, signaling deconfinement.}
\label{Fig1}
\end{figure}

\begin{figure}
\includegraphics[width=8cm]{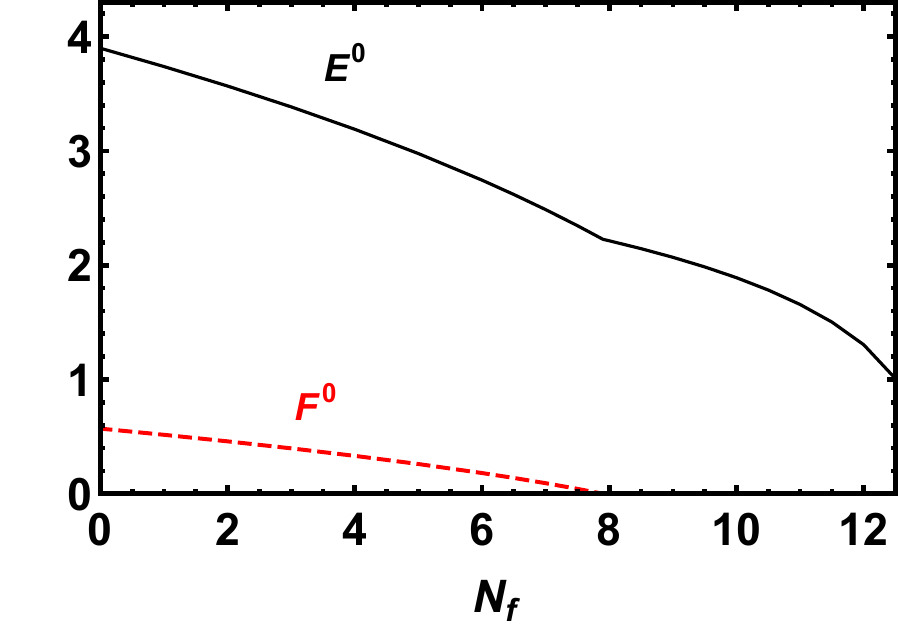}
\caption{The behavior of the $N_f$-dependent invariant BSAs $E^{0}$ and $F^{0}$ in the chiral limit is examined. While $F^{0}$ approaches zero at the chiral symmetry restoration point $N^{c}_{f}$, $E^{0}$ retains non-zero values in this region, normalized around $N_f = 12.5$ as a function of $N_f$.
 }
\label{Fig2}
\end{figure}

\begin{figure}
\includegraphics[width=8cm]{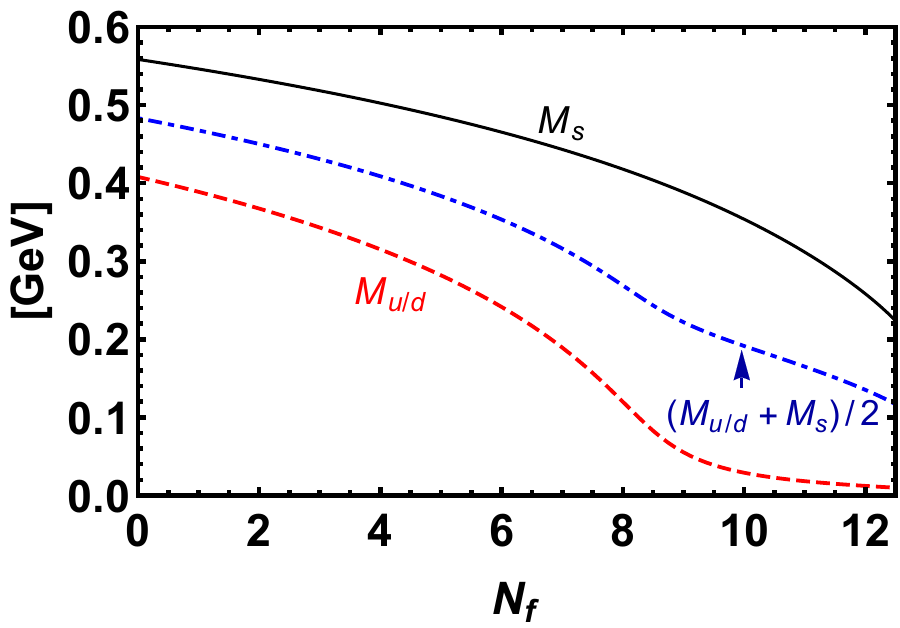}
\caption{Behavior of the dressed quark mass $M_{u/d}$ (with $m_{u=d}=0.007$) GeV, and dressed  strange quark mass $M_s$ (with bare mass $m_s=0.17$ GeV), and the average dressed mass  $(M_{u/d}+M_{s})/2$, for various number $N_f$. All the parameters monotonically decrease as we increase $N_f$.
}
\label{Fig3}
\end{figure}

\begin{figure}
\includegraphics[width=8cm]{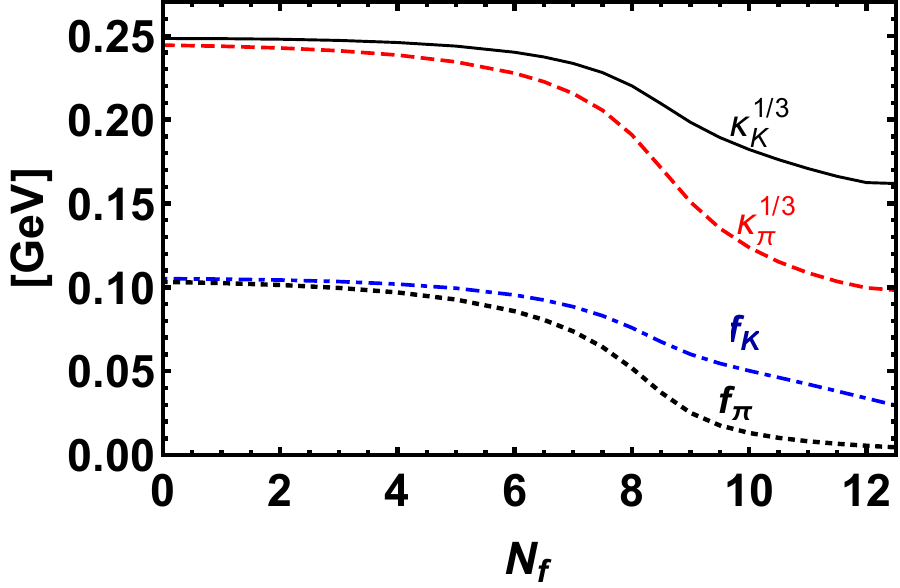}
\caption{The behavior of the  leptonic pion decay constant $f_{\pi}$, the kaon decay constant $f_K$,  the in-pion condensate  $\kappa^{1/3}_{\pi}$ and  the in-kaon  condensate $\kappa^{1/3}_{K}$ as a function  $N_f$ is analyzed  with bare quark masses. All the parameters monotonically decrease as $N_f$ increases, indicting a partial restoration of  dynamical  chiral symmetry  above $N^{c}_{f}\approx 8.2$.}
\label{Fig4}
\end{figure}

\begin{figure}
\includegraphics[width=8cm]{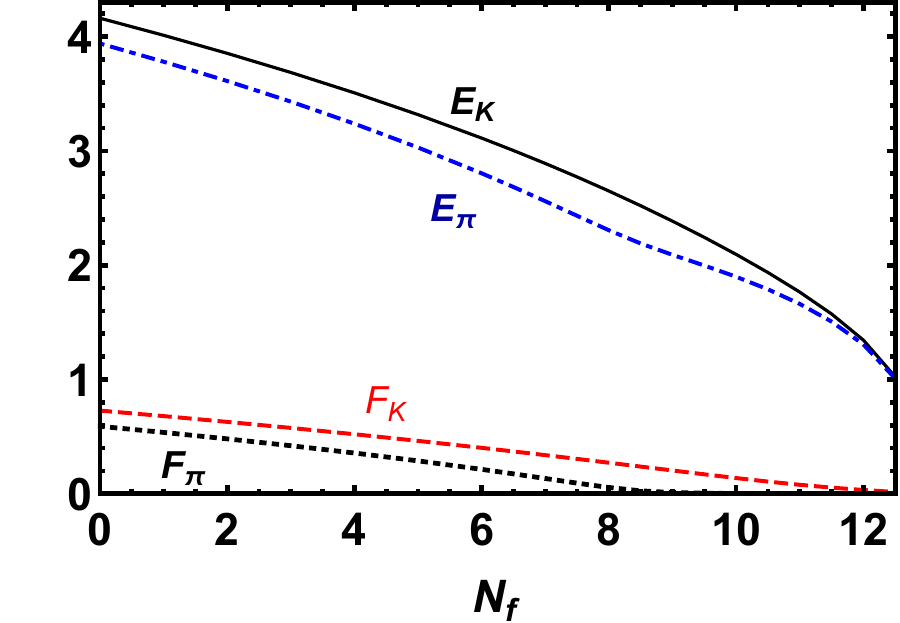}
\caption{Behavior of the  invariant BSAs for pion $E_{\pi}$, $F_{\pi}$ and kaon $E_{K}$, $F_{K}$ , with bare quark masses as a function of $N_f$.}
\label{Fig5}
\end{figure}

\begin{figure}
\includegraphics[width=8cm]{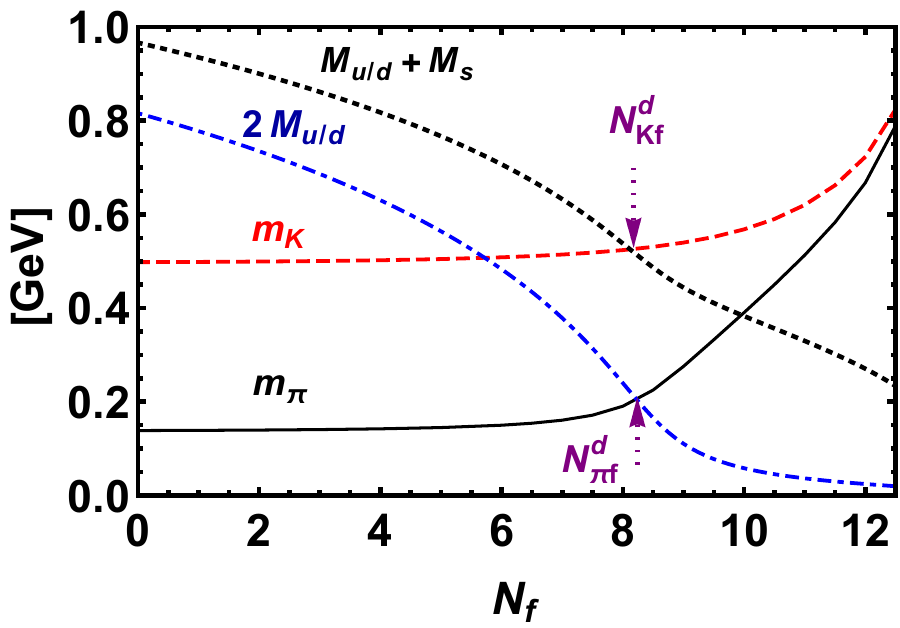}
\caption{
The behavior of the pion and kaon masses, the sum of the light dressed quark masses, and the sum of the light dressed quark mass and the strange quark masses are illustrated in this plot. It shows that the pion mass $m_\pi$  intersects  $2M_{u/d}$ at the critical flavor dissociation number $N^{d}_{\pi}$, while the kaon mass $m_K$ intersects $M_{u/d} + M_s$ at $N^{d}_{f,k}$. Notably, at $N^{d}_{f,k} \approx 8.2$, both the pion and kaon dissociate into their constituent quarks.}
\label{Fig6}
\end{figure}

\begin{figure}
\includegraphics[width=8cm]{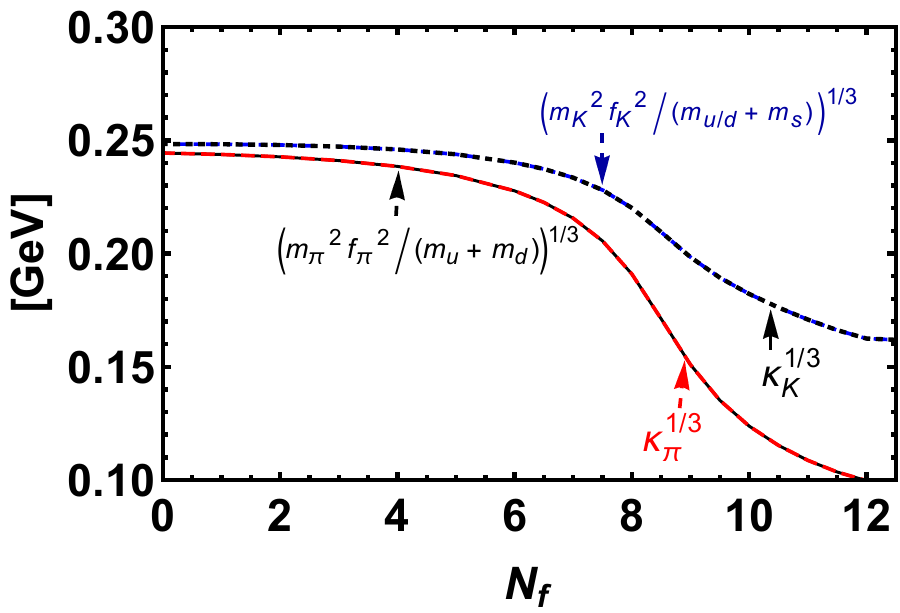}
\caption{The verification of the GMOR relation as a function of $N_f$ for both in-pion and in-kaon condensates is presented. This plot demonstrates that the in-pion and in-kaon condensates satisfy the corresponding values derived from the GMOR relation for various numbers of $N_f$. 
}
\label{Fig7}
\end{figure}

In this section, we present our numerical findings and discussions. We begin by numerically solving the SDE to calculate the dressed quark mass in the chiral limit($m_{f}=0$) and with the bare quark masses $m_{u/d}$ and $m_s$, i.e.,  $M_{u/d}$-and $M_s$, using the FCI model, while varying the  number of flavors $N_f$.  Next, we solve the homogeneous Bethe-Salpeter equation to determine the properties of pseudoscalar meson bound states in the chiral limit  and with the bare quark masses. We primarily focus on the properties of pseudoscalar meson bound states in their ground states, including the mass of the bound state $m_{0^{-+}}$, the leptonic pion decay constant $f_{0^{-+}}$, the in-pseudoscalar meson condensate $\kappa^{1/3}_{0^{-+}}$, and the invariant BSA $E_{0^{-+}}$ and $F_{0^{-+}}$ as a functions of $N_f$. The values of the  parameters to be used in the FCI model are enlisted in Table~\ref{tab1}. 
\begin{table}
\caption{\label{tab1} Parameters of the FCI model used as inputs for the SDE  and BSE. These parameters  were determined to reproduced the properties of  ground state $\pi$-and $\rho$-mesons (for $N_f=2$ and $N_c=3$)
in Ref.~\cite{Roberts:2010rn} and  for $K$-meson in~\cite{Chen:2012txa}, except  an additional parameter in FCI model which was determined in~\cite{Ahmad:2020jzn}. All the enlisted dimensioned quantities are in GeV.}
\begin{tabular}{p{1cm} p{1cm} p{1cm} p{1cm} p{1cm} p{1cm} p{1cm}}
\hline
$m_{u=d}$ & $m_{s}$ & $\Lambda_{ir}$& $\Lambda_{uv}$ & $\alpha_{ir}$& $m_{g}$ & $ \mathcal{N}^{c}_{f} $\\
\hline
$0.007$   & $0.17$   & $0.24$ & $0.905$ & $0.93 \pi$ & $0.8$ & $10.9$\\
\hline
\hline
\end{tabular}
\end{table}
To ensure the consistency of our results, we first set $N_f=2$ and $N_c=3$. We derived the properties of the Goldstone bososn, $\pi$ and $K$ mesons using the FCI Model based on the SDE-BSE equations, and we have summarized these results in Table~\ref{tab2}. In this context, our findings from the FCI model align well with those obtained from the CI model.~\cite{Roberts:2010rn,Chen:2012qr, Chen:2012txa,Xing:2022jtt,Hernandez-Pinto:2023yin}.
\begin{table*}[t]

\caption{\label{tab2} In the following table we presents the  solution of the SDE and the BSE and calculated  all these physical quantities for fixed $N_f=2$. All the computed quantities are in the units GeV.}  
\begin{tabular}{p{0.4cm} p{0.4cm} p{0.4cm} p{0.4cm} p{0.4cm} p{0.4cm} p{0.4cm} p{0.4cm} p{0.4cm} p{0.4cm} p{0.4cm} p{0.4cm} p{0.4cm} p{0.4cm} p{0.4cm} p{0.4cm}}
\hline
$M_{0}$& $M_{u/d}$&$M_{s}$& $E_{0}$& $F_{0}$& $E_{\pi}$& $F_{\pi}$& $E_{K}$& $F_{K}$& $f_{0}$& $f_{\pi}$& $f_{k}$& $\kappa^{1/3}_{\pi}$ &$\kappa^{1/3}_{K}$& $m_{\pi}$&$m_{k}$\\ 
\hline
0.58& 0.368 & 0.533 & 3.568& 0.459&3.61 & 0.48 &3.86&0.62 & 0.1& 0.101& 0.11 & 0.243& 0.248&0.1395&0.499\\
\hline
\hline
  \end{tabular}
  \label{tab:2}
\end{table*}
 In the chiral limit, we plotted various quantities in Fig.~\ref{Fig1}, including the dressed quark mass $M_{0}$, the quark-antiquark condensate $(-<\bar{q}q>^{1/3}_{0})$, the leptonic decay constant $f_{0^{-+}}(=f_{0})$, the mass of the bound state Goldstone boson $m_{GB}$, and the confinement scale $\tilde{\tau}_{ir}$  (shown in the sub-figure within Fig.~\ref{Fig1}) as a function of $N_f$. Notably, $M_{0}$, ($-<\bar{q}q>^{1/3}_{0}$) and the $f_{0}$ vanish at and above $N_f=N^{f}_{c}\approx8$, where dynamical chiral symmetry is restored. At and above $N^{c}_{f}$, the confinement scale $\tilde{\tau}_{ir}$ approaches infinity (or equivalently, $\Lambda_{ir}\rightarrow0$~\cite{ahmad2025qcdphasediagramlarge}). This divergence is consistent with the selection in Eq. ~(\ref{CI10}), which indicates that quark production thresholds reemerge, resulting in deconfinement at approximately $N^{c}_{f}\approx 8$. Just below this critical number $N^{c}_{f}$, the value of  $\tilde{\tau}_{ir}$ is approximately $113$ GeV$^{-1}$, while at and beyond $N^{c}_{f}$, it becomes infinite, as anticipated from the choice in Eq.~(\ref{CI10}).\\ 
The mass of the bound state Goldstone boson, denoted as $ m_{GB}$, remains massless in the region where dynamical chiral symmetry is broken, until it reaches approximately $N_f = N^{d}_{f} \approx 8$. At this point, the mass suddenly increases. Physically, this transition indicates that the massless Goldstone boson separates from a bound state, transforming into resonant particles. This occurs as quark-antiquark pairs are released from the bound state, signaling a deconfinement phase (similar to what happens in a heat bath \cite{Hufner:1996pq}) at the critical flavor  dissociation point $ N^{d}_{f} \approx 8 $, which coincides with $N^{c}_{f}$.
\\
In analogy to the heat bath~\footnote{(i.e.,
At finite temperature $T$, the dissociation of bound states  at Mott temperature $T_M$ is in some sense closely related to the deconfinement transition
~\cite{Hufner:1996pq}, where the critical
temperature $T_c$ of the deconfinement transition of QCD separates two phases: bound
states (baryons and mesons) for $T < T_c$ from a plasma of constituents (quarks and
gluons), for $T > Tc$. The Mott transition temperature $T_M $, on the other hand, separates a
phase consisting of bound states and constituents for $T < T_M $ from a phase of constituents
only at $T > T_M $~\cite{Hufner:1996pq}. Thus the Mott and deconfinement transitions have in common that their
bound states delocalize into their constituents~\cite{Hufner:1996pq})}
, in our flavor dependent version, the critical number of flavor $N^{f}_{c}$ and the confinement scale $\tilde{\tau}_{ir}$, separate the bound
states at $N_f < N^{c}_{f}$  from a  constituents (quarks and
gluons) at $N_f > N^{c}_{f}$ . On the other hand, the dissociation critical number of flavors $N^{d}_{f}$ plays the role of Mott temperature $T_M$~\cite{Hufner:1996pq}, separates a
phase consisting of bound states and constituents for $N_f< N^{d}_{f}$ from a phase of constituents
only at $N_f > N^{d}_{f}$.  In  Fig.~\ref{Fig2}, we present the  invariant BSA $E^{0}$ (or $E_{0^{-+}}^{0}$ ) and $F^{0}$ ($F^{0}_{0^{-+}})$. The $F^{0}$ vanishes in the chiral symmetry restoration region around $N^{c}_{f}$. In contrast, $E^{0}$ has a no-zero value at $N^{c}_{f}$ but shows a slight kink at this point, ultimately reaching unity at  $N_f=12.5$.  
Next, we focus on the case beyond the chiral limit, where ($m_f\neq0$). 
longer exact, It is widely accepted that once a small bare quark mass is considered, chiral symmetry is no longer exact and the pion becomes slightly massive according to Gell-Mann-Oakes-Renner relation (GMOR). We use a bare quark mass of  $m_{u,d}=0.007$ GeV and $m_{s}=0.17$ GeV to obtain the dressed mass $M_{u/d}$,  the dressed strange quark mass $M_s$ and their average $(M_{u/d}+M_s)/2$ as a function of $N_f$, as  shown in Fig.~\ref{Fig3}. All quantities exhibit a monotonically decreasing trend with increasing $N_f$, but they do not vanish in the asymptotic limit because chiral symmetry is explicitly broken by the finite bare quark mass. In this scenario, chiral symmetry is not fully restored but is partially restored; the dynamically generated dressed mass approaches zero, leaving only the bare mass, such that ($M_f\rightarrow m_f$).
The in-pion condensate $\kappa^{1/3}_{\pi}$, the in-kaon condensate $\kappa^{1/3}_{K}$, the leptonic decay constant for the pion $f_{\pi}$, and for the kaon $f_{K}$  as functions of $N_f$ are illustrated in Fig.~\ref{Fig4}. The critical number of flavors in this context can be derived from the flavor-gradient of the corresponding in-meson condensate, yielding $N^{c}_{f}\approx 8.2$ . Above this threshold, chiral symmetry is partially restored.
In Fig. \ref{Fig5}, we present the invariant BSAs for the pion, $E_{\pi}$, $F_{\pi}$, and for the kaon, $E_{K}$ and $F_{K}$. These BSAs exhibit a decreasing trend as the number of flavors, $N_f$, increases. Notably, the invariant amplitudes $F_{\pi}$ and $F_{K}$ remain non-zero in this scenario due to the use of bare mass. Additionally, the amplitudes $E_{\pi}$ and $E_{K}$ approach unity as $N_f$ becomes sufficiently large.
In Fig.~\ref{Fig6}, we illustrate the behavior of the masses of the bound state pion $ m_{\pi}$, kaon $ m_{K}$, $2M_{u/d}$, and $M_{u/d} + M_s$ as a function of the number of flavors $N_f$. As previously mentioned, in the chiral limit, a phase transition to a state of chiral symmetry restoration occurs at a critical flavor number $N^{c}_{f}$. Simultaneously, $N^{d}_{f}$ marks the point at which the Goldstone boson transitions from a bound state to a dissociated state. Building on our discussion of the chiral limit, we further examined the dissociation of the bound state at the point $N^{d}_{f}$.
In the case of pions, there exists a dissociation point defined by the condition $2 M_q(N^{d}_{f,\pi}) = m_{\pi} (N^{d}_{f,\pi})$, analogous to the concept of a heat bath~\cite{Hufner:1996pq}. Similarly, we define a kaonic flavor dissociation number $N^{d}_{f,K}$, at which the relationship $M_{u/d}(N^{d}_{f,K})+ M_s(N^{d}_{f}) = m_K(N^{d}_{Kf})$ holds. In Fig.~\ref{Fig6}, we plot the bound state masses of the pion $m_{\pi}$, kaon  $m_{K}$, $2M_{u/d}$, and $M_{u/d}+M_s$ as functions of $N_f$
This plot illustrates that $2M_{u/d}$ decreases monotonically as $N_f$ increases. In the region where chiral symmetry is broken, the pion mass $m_{\pi}$ and kaon mass $m_{K}$  gradually increase. However, in the chiral symmetry restoration region, these masses rise sharply. We identified the intersection point at  $N^{d}_{\pi f}$, where the pion mass  $m_{\pi}$ intersects $2M_{u/d}$, while the kaon mass $m_{K}$ intersects with $M_{u/d}+M_s $ at $N^{d}_{Kf}$. Notably, the critical flavor number for pion and kaon dissociation coincides, yielding $N^{d}_{K f}=N^{d}_{\pi f}\approx 8.2$. Therefore, the critical number of flavors associated with dissociation defines a state beyond which bound states no longer exist. This is equivalent to stating that this $N^{d}_{(\pi,K )f}$ is the point at which the mass of the bound state equals the mass of its constituent particles.  Thus, our
results  suggest  that  the  chiral  and  deconfinement  phase
transitions occur  at  the  same critical number of flavors $N^{c}_{f}=N^{d}_{f}$, in the chiral limit as well as in the presents of bare quark masses $N^{c}_{f}=N^{d}_{(\pi,K)f}$.
We verify the consistency of our results by examining the GMOR relation, as illustrated in Fig.~\ref{Fig7}. By substituting the values of $f_{\pi}$, $m_{\pi}$, $f_{K}$, and $m_{K}$ derived from solving the BSE into the GMOR relations, specifically Eq.~(\ref{BSE8g}) and Eq.~(\ref{BSE8h}), we calculate the right-hand side of the GMOR relations. We then compare these values with the in-pion and in-kaon condensates obtained from Eq.~(\ref{BSE8c}). Our findings indicate that our results are consistent with the GMOR relation across all considered flavors and exhibit a monotonic decrease with increasing $N_f$. In the next section, we will summarize our findings and future perspectives.


\section{Summary and perspectives} \label{section 5}
In this research paper, we explore the properties of pseudoscalar Goldstone bosons, specifically the massless pion in the chiral limit, as well as the $\pi$ and $K$ mesons. Our focus includes their dressed masses, in-pseudoscalar meson condensates, and leptonic decay constants across various values of $N_f$. We employed the FCI model and utilized the SDE approach to calculate the dressed quark masses in the chiral limit, denoted as $M_0$, along with the bare quark masses $M_{u/d}$ for up or down quarks (with $m_{u} = m_{d} = 0.007$ GeV to ensure isospin symmetry) and the dressed mass of the strange quark $M_s$ (with $m_s = 0.17$ GeV), taking into account different values of $N_f$. Using the BSE, we have computed various properties of bound state pseudoscalar mesons in both the chiral limit and with bare quark masses, for different values of $N_f$. To verify our findings, we initially focused on the properties of the Goldstone Boson in the chiral limit and with bare masses, specifically examining the $\pi$ and $K$-mesons with a fixed $N_f=2$. Our results are consistent with previous model calculations based on the SDE-BSE approach and align with the GMOR relation. Following this, we varied $N_f$ and predicted the behaviors of all observed quantities as outlined below: \\
1) In the chiral limit, as the number of flavors $N_f$ increases, dynamical chiral symmetry is restored, leading to deconfinement at a critical number of  flavors, $N^{c}_{f} \approx 8$. At this threshold, some parameters vanish, the confinement scale $\tilde{\tau}_{ir}$ diverges. However, for $N_{f} > N^{c}_{f}$ the mass parameter of the Goldstone boson $m_{GB}$ remains  constant in the chiral symmetry broken phase until $N_{f}> N^{c}_{f}$, where the dynamical chiral symmetry is fully restored, $m_{GB}$ sharply rises.  This transition marks a shift of the Goldstone boson from a bound state to a resonant particle, ultimately leading to dissociation at $N^{d}_{f}$ (the bound state dissociation  critical number of flavors), which aligns with $N^{c}_{f}\approx8$. This dissociation signifies deconfinement, allowing quarks and antiquarks to escape from their bound states. Thus, $N^{d}_{f}$ is associated with a state is defined to be the number of flavors above which
the bound state no longer exists. Equivalently this is
the critical number of flavors, at which the mass of the
state becomes equal to the mass of its constituents. \\
2) When the bare quark masses are considered, all the parameters under observation except for the bound state masses  $m_{\pi}$ and $m_{K}$, monotonically decreases as we increase $N_f$. As in this case, the chiral symmetry is explicitly broken and partially restored at  and above $N^{c}_{f} \approx 8.2$. The  $m_{(\pi, K)}$  increases  quickly and dissociate at and above  critical dissociation number of flavors $N^{d}_{f} =N^{d}_{(\pi, K)f}$, this $N^{d}_{(\pi, k)f}$ separates the bound states of the pion and kaon from  their constituents. In the case of pions, there exists a dissociation point defined by the equation $2 M_{u/d}(N^{d}_{\pi f}) = m_{\pi} (N^{d}_{\pi f})$. Additionally, for kaons,  $N^{d}_{Kf}$ can be identified as, $M_{u/d}(N^{d}_{K f}) + M_s(N^{d}_{Kf}) = m_K(N^{d}_{Kf})$, similar to the concept of a heat bath~\cite{Hufner:1996pq}. Interestingly, the critical flavor dissociation numbers for both pions and kaons coincide, yielding $N^{d}_{K f} = N^{d}_{\pi f} \approx 8.2)$. \\
 3) The  bound state dissociation number of flavors $N^{d}_{(\pi,K)f}$ viewed as the critical number of flavors necessary for both chiral symmetry restoration and deconfinement, similar to the Mott temperature observed in finite temperature QCD~\cite{Hufner:1996pq,Xu:2021lxa}).\\
 4) The findings for a fixed number of flavors $N_f = 2$, align well with prior predictions regarding the properties of pions and kaons, which were based on SDE-BSE analyses and experimental data. .\\
5) Additionally, our predictions from the FCI model for various $N_f$ counts have successfully met the GMOR relation.\\
In summary, the relationship between the number of flavors $N_f$ and chiral symmetry is essential in understanding the dynamics of confinement and deconfinement. This relationship underscores a critical threshold that significantly influences particle behavior in the system, potentially having a more pronounced effect on light hadron phenomenology.
Looking ahead, we aim to expand this research to compute the electromagnetic form factors of the pion and kaon, as well as the scalar and vector meson bound states, and explore other related phenomena for higher values of $N_f $.

\bmhead{Acknowledgements}
A. Bashir, B.Masud and M.A. Bedolla for their guidance and suggestion. We also
grateful to the colleagues of  the  Institute of Physics,  Gomal University for their hospitality and
support. 

\section*{Declarations}
\begin{itemize}
\item Funding: No funding received for this work.
\item  We have no conflict of interest/competing interests.
\item Data availability :
This work has no associated data. This is theoretical study and no experimental data included.
\item Author contribution; Aftab Ahmad  wrote the manuscript 
\end{itemize}








\bibliography{sn-bibliography}

\end{document}